\documentclass[acmsmall, nonacm]{acmart}
\renewcommand\footnotetextcopyrightpermission[1]{}

\AtBeginDocument{%
  }

\setcopyright{acmlicensed}
\copyrightyear{2025}
\acmYear{2025}
\acmDOI{XYZ}

\setcopyright{none}
\acmDOI{}
\acmISBN{}
\acmConference{}
\acmYear{}
\acmPrice{}

\acmConference[AISC'2025]{Australasian Information Security Conference}{February 10--14,
  2025}{Brisbane, Australia}
\begin{document}

\title{Predicting IoT Device Vulnerability Fix Times with Survival and Failure Time Models}

\author{Carlos A. Rivera A.}
\email{c.riveraalvarez@unsw.edu.au)}
\orcid{1234-5678-9012}
\author{Xinzhang Chen}
\email{xinzhang.chen@unsw.edu.au}
\author{Arash Shaghaghi}
\email{a.shaghaghi@unsw.edu.au}
\author{Gustavo Batista}
\email{g.batista@unsw.edu.au}
\author{Salil S. Kanhere}
\email{salil.kanhere@unsw.edu.au}
\affiliation{%
  \institution{School of Computer Science and Engineering, The University of New South Wales (UNSW Sydney)}
  \city{Sydney}
  \state{NSW}
  \country{Australia}
}

\renewcommand{\shortauthors}{Rivera A. et al.}

\begin{abstract}
 % In this paper, we propose the time to fix framework. The proposed framework helps predict the time it takes for a patch to be released for a vulnerable Internet of Things (IoT) device, allowing system administrators to manage better risks associated with potential vulnerabilities. The time to fix framework is based on the risk level associated with each vulnerability and the predicted time to patch the device. Our proposal utilises a constructed dataset with IoT devices, associated vulnerabilities, Twitter trends and accelerated failure time (AFT) survival analysis over the assembly regression XGBoosting model. Our time to fix framework boasts numerous benefits. Firstly, it calculates the time needed to fix any vulnerabilities in IoT devices. Secondly, it liberates system administrators from relying on uncertain responses from manufacturers by automatically providing them with time to fix information for their devices. Lastly, we have utilised AFT survival analysis and the ensemble XGBoost model to ensure precise prediction of the likelihood of time to fix estimates.
The rapid integration of Internet of Things (IoT) devices into enterprise environments presents significant security challenges. Many IoT devices are released to the market with minimal security measures, often harbouring an average of 25 vulnerabilities per device. To enhance cybersecurity measures and aid system administrators in managing IoT patches more effectively, we propose an innovative framework that predicts the time it will take for a vulnerable IoT device to receive a fix or patch. We developed a survival analysis model based on the Accelerated Failure Time (AFT) approach, implemented using the XGBoost ensemble regression model, to predict when vulnerable IoT devices will receive fixes or patches. By constructing a comprehensive IoT vulnerabilities database that combines public and private sources, we provide insights into affected devices, vulnerability detection dates, published CVEs, patch release dates, and associated Twitter activity trends. We conducted thorough experiments evaluating different combinations of features, including fundamental device and vulnerability data, National Vulnerability Database (NVD) information such as CVE, CWE, and CVSS scores, transformed textual descriptions into sentence vectors, and the frequency of Twitter trends related to CVEs. Our experiments demonstrate that the proposed model accurately predicts the time to fix for IoT vulnerabilities, with data from VulDB and NVD proving particularly effective. Incorporating Twitter trend data offered minimal additional benefit. This framework provides a practical tool for organisations to anticipate vulnerability resolutions, improve IoT patch management, and strengthen their cybersecurity posture against potential threats.
 
\end{abstract}


\received{18 November 2024}
\received[revised]{}
\received[accepted]{}

\maketitle

\section{Introduction}

As of 2022, enterprises have deployed over 100 million Internet of Things (IoT) devices, projected to surpass 8 billion by 2030 \cite{rondon2022survey, VailsherySujay082022}. This rapid integration underscores a significant interest in adopting IoT technologies within organisational settings. However, the swift expansion of IoT devices amplifies security concerns, increasing potential attack surfaces and complicating the management of device vulnerabilities. Two risks of particular interest to this research are difficulties in validating device security \cite{carlos, MadelineLauver2022, JeffCostlow2022} and implementing timely patch management in organisations \cite{FBIPIN2022}.

The absence of robust standards and legal frameworks often leaves device security at manufacturers' discretion, who may prioritise cost, size, usability, and time-to-market over security considerations \cite{maroof2019plar}. Consequently, IoT devices are frequently released without adequate security measures, leading to an average of 25 potential vulnerabilities per device \cite{lee2015internet}. Assessing and mitigating these vulnerabilities is costly and time-consuming, especially as the number and variety of devices continue to grow.

While large corporations typically employ cost-benefit analyses to guide technological investments, they often overlook the financial risks associated with device vulnerabilities and threats \cite{lee2020internet}. This oversight can result in deploying insecure devices, which may become targets for exploitation, as evidenced by global security incidents like the Mirai botnet \cite{antonakakis2017understanding}. Effective patch management is crucial for mitigating security risks associated with IoT devices, ensuring that vulnerabilities are addressed promptly before malicious actors can exploit them. Failure to do so increases the risk of cyber-attacks and can lead to significant financial and reputational damage.

%Despite awareness of these risks, vulnerable devices frequently remain connected without receiving necessary patches, exacerbating security threats.

%Recent cases, such as the security weaknesses found in Peloton fitness equipment, demonstrate that individuals and organizations continue to face persistent dangers from unpatched IoT devices. Manufacturers may downplay vulnerabilities by asserting that exploitation requires physical access; however, until these vulnerabilities are addressed and patches are applied, the devices remain insecure. This situation underscores the critical need for effective vulnerability management and timely patching in IoT devices.

We introduce an innovative framework to predict when a vulnerable IoT device will receive a patch. Extending beyond traditional patch management techniques \cite{li2017large}, our framework enables system administrators to anticipate patch release timelines, thereby enhancing IoT device management and strengthening cybersecurity defences.

This paper offers three key contributions:

\begin{itemize}
\item \textbf{Time-to-Fix Prediction Framework}: We develop a predictive model that forecasts the duration until vulnerable IoT devices receive fixes or patches. Our framework provides accurate time-to-fix predictions using a survival analysis approach based on accelerated failure time, implemented with the XGBoost ensemble regression model.

\item \textbf{Comprehensive IoT Vulnerabilities Database}: We compile a rich database combining information from public and private sources, offering detailed insights into affected devices, vulnerability detection dates, published CVEs, and patch release dates. The dataset includes real-time analysis features, such as trending Twitter activity related to CVEs. We share publicly the dataset on a GitHub\footnote{https://github.com/criveraalvarez/Time2FixPredictionDS/} page.  

\item \textbf{Feature Evaluation}: We conduct testing to assess the effectiveness of various feature combinations in our model. These features encompass (1) fundamental data on devices and vulnerabilities; (2) vulnerability information from the National Vulnerability Database (NVD), including CVE identifiers, Common Weakness Enumeration (CWE), and Common Vulnerability Scoring System (CVSS) scores; (3) textual descriptions of vulnerabilities transformed into sentence vectors; and (4) frequency analysis of CVE-related Twitter trends.
\end{itemize}

The remainder of this paper is structured as follows: Section \ref{litrev} reviews the background and related work relevant to our research. Section \ref{mainsection} details our dataset's formulation, our predictive framework's design, and the evaluation metrics for our Time-to-Fix model. Section \ref{ExperimentsandEvaluationResults} presents and discusses our experimental results. Finally, Section \ref{Conclusion} concludes the paper with a discussion of our findings and their implications for future research.

\section{Related Work}\label{litrev}

\subsection{Time to Exploit Prediction}\label{time2exploit}
{Exploits are software programs purposely developed to take advantage of any identified weaknesses or vulnerabilities in a system. Exploits are directly linked to patches since the first can be prevented with the latter's release. Using this scenario, researchers in \cite{leverett2022vulnerability} researched the potential number of vulnerabilities that could emerge within a system over a specific timeframe. The researchers' primary objective was to anticipate vulnerabilities before the occurrence of any exploits. The study highlighted that 80\% of exploits typically manifest themselves 23 days before the publication of the related Common Vulnerabilities and Exposures (CVE). The increasing pressure exerted by exploit creators has compelled organisations to intensify their efforts to release patches to address vulnerabilities before disclosing them publicly. Consequently, this has posed a challenge for researchers in anticipating vulnerabilities and exploits' appearance.} 

{In a recent study by Bhatt et al. \cite{bhatt2021exploitability}, the authors addressed the challenge of predicting software exploits by proposing a framework that employs a Decision Tree binary classifier algorithm. To determine the most suitable algorithm, the framework was compared with other classifiers such as Random Forest, Naive Bayes, Logistic Regression, and Supervised-vector Machines. The training data for these algorithms consisted of records from the National Vulnerability Database (NVD) and Exploits databases. However, one limitation of the study was the failure to incorporate the timing of exploit appearance. In our implementation, we aim to address this limitation by predicting the time at which a fix for the vulnerability would appear rather than solely focusing on predicting the time of exploit appearance. 

Tavabi et al. \cite{tavabi2018darkembed} explored the prediction of software vulnerability exploitation. The study involved crawling the dark and deep web to amass a dataset of approximately 2.5 million records containing references to CVEs and associated exploit availability. This data underwent compilation, filtration, and processing through three embedding algorithms, yielding contextual linkages between the information and the CVEs and exploits. The resultant embeddings were then vectorised and integrated with supplementary features sourced from NVD, ExploitDB, Metasploit, and Symantec's antivirus database. This comprehensive methodology facilitated the authors' ability to predict the presence of a software vulnerability exploit. This study does not address the prediction of exploit timing. }

{In a study conducted by Almukainizi et al. \cite{almukaynizi2017proactive}, a predictive model was proposed to determine the likelihood of a vulnerability being exploited based on executed a two-fold analysis approach, encompassing binary-based and text-based data. The binary data was converted into a binary format. In contrast, the text data underwent further analysis using text frequency on a bag-of-words approach to transform it into numerical values. The resulting dataset included the analysed features, Common Vulnerability Scoring System (CVSS) scores, language information, and mentions in online forums. The primary sources of the data were the National Vulnerability Database (NVD) and Exploit Database (ExploitDB), along with forum mentions extracted from the deep web and dark web. To predict the likelihood of exploitation, the study evaluated several machine learning models, including Random Forest, Support Vector Machine, Naive Bayes, and Logistic Regression. The Random Forest algorithm was ultimately selected due to its superior performance in terms of the F1 measure. It is worth noting, however, that while the proposed model can predict the potential exploitability of a vulnerability, it does not provide information regarding the timing of the exploit's occurrence. Furthermore, the approach employed in the prediction phase, which involved amalgamating features based on their respective sources, proved instrumental. Multiple predictions were made using this approach, which was effective in our study. The inclusion of features associated with specific sources provided us with the ability to ascertain the significance of certain features and their sources.}

{In the work by Jacobs et al., the authors presented two publications discussing innovative concepts and valuable guidance that are worth replicating. In their initial publication \cite{jacobs2021exploit}, the researchers proposed the creation of an extensive dataset encompassing Common Vulnerabilities and Exposures (CVEs) and associated exploits sourced from various outlets. This dataset was segmented into nine sub-datasets, each corresponding to its source. The data was leveraged to train an Extreme Gradient Boosting Trees model. The methodology employed to evaluate the model and the dataset involved various combinations of data from the sub-datasets, culminating in the complete model that integrated all the data. Notably, the complete dataset comprised 209 features linked to 75585 distinct CVEs. These endeavours were undertaken to predict the Common Vulnerability Scoring System (CVSS) and determine the presence of associated exploits for the CVEs. In the latter publication \cite{jacobs2020improving}, the authors created an exploit score to predict the probability of a vulnerability being exploited in the next 12 months after the vulnerability is released publicly. The authors' analysis of vulnerabilities using time as a factor needs to be revised to account for the various factors that can influence a vulnerability's lifespan. Neither proposal specifically predicted the time frame within which the exploits would manifest.}

{The study by Chen et al. \cite{chen2019vest} aimed to forecast the probability of exploiting a vulnerability and the anticipated timeframe for such occurrences. The methodology involved amalgamating data sourced from the public platform Twitter with exploitation signatures and specific vulnerability details. While the proposed approach successfully provided estimations for the appearance of an exploit linked to a Common Vulnerabilities and Exposures (CVE) entry, the study lacked empirical evidence or implementation specifics. 

{The significance of Twitter and vulnerability-related data in predicting the emergence of exploits has been acknowledged by Sabottke et al. \cite{sabottke2015vulnerability}. The authors collected data from various sources, including Twitter, ExploitDB, OSVDB, Microsoft Security Advisories, and Symantec's antivirus and intrusion protection signatures. They employed a Supervised-vector Machine algorithm to predict the timing of exploit appearance, using diverse feature combinations from multiple sources to enhance prediction accuracy. An important element of the study involved correlation analysis, which juxtaposed exploit appearance times with tweet occurrences, determining Twitter's value as an information source. While the study focused on predicting the date of the vulnerability exploit appearance, it does not consider the vulnerability patch release time.}
% Such information would significantly aid in the formulation of more effective management strategies.

{In their study Chen et al. \cite{chen2019using}, the authors utilized Twitter data to forecast the timing of vulnerability exploitations. They curated a comprehensive dataset by amalgamating information from Twitter, CVEs from NVD, and Intrusion Protection Signatures from Symantec. Employing machine learning algorithms such as Support Vector Machines (SVM), Logistic Regression (LR), Naive Bayes (NB), Random Forest (RF), and XGBoost, the authors aimed to predict the timeframe within which a vulnerability would be exploited. While the methodology employed in this study demonstrated high fidelity with ground truth, the authors also suggested that Twitter could potentially serve as the primary source for detecting exploit releases.}
%However, our research indicates that while Twitter is a valuable resource, it is not indispensable in this context.

{Due et al. \cite{du2023expseeker} integrated Twitter datastream and CVSS data into their exploit prediction model. Findings underscore the importance of leveraging information from explicit vulnerability disclosures in the NVD and Twitter posts. A distinguishing feature of this study is the utilisation of text embeddings from Twitter to generate vectorized features that are then fed into a Bidirectional Long Short-Term Memory (Bi-LSTM) model. Notably, the model incorporates a Conditional Random Field (CRF) module, which assigns sequences of tags to each word. The methodology relies solely on Twitter as a data source, with validation from the Alexa top one million and Virus Total API. While the model's approach is intriguing, the results from our study indicate that its utility is somewhat limited, suggesting that alternative data sources could be used, as Twitter's future availability and operational format are uncertain. Additionally, while this implementation predicts whether a Common Vulnerability and Exposure (CVE) would be associated with the appearance of an exploit, it does not offer predictions for the estimated appearance date.}

%The authors of \cite{suciu2022expected} introduce a novel exploitability metric, suggesting that it could be more efficient and accurate than existing alternatives. However, their approach misused CWE IDs for categorizing exploit types, leading to inaccuracies. Additionally, the features extracted from Twitter were analyzed using frequency analysis techniques, which overlooked valuable information such as trend counts. This work adopts a classification-based approach. Other methods can be used to predict a time-based value. For instance, in \cite{householder2020historical}, the authors utilised Cox proportional hazard models from the survival analysis approach. 

\subsection{Survival Analysis}\label{survival}
{Survival analysis is an essential sub-area of statistical modelling, and it offers a variety of methods to deal with censored data problems. This issue arises when the event of interest cannot be observed due to time constraints or when the event time is lost. This type of analysis, called time-to-event data, is designed to model a specific event of interest (In our case, the event is represented by whether a vulnerability has been patched or not) and estimate the time for such an event (time to fix). In health science, survival analysis has found extensive applications, particularly in scenarios where time is crucial in determining outcomes such as recovery, survival, or mortality. 

In a study conducted by Chai et al. \cite{chai2017new}, the researchers undertook the combination of two prominent survival analysis methodologies, namely the Cox proportional hazards model (Cox) and the Accelerated Failure Time model (AFT). Their investigation revealed that both approaches demonstrated efficacy in predicting outcomes. Specifically, the Cox model was employed to classify cancer patients, while the AFT model was utilized for predicting censored data. Ultimately, integrating these models into a combined COX-AFT framework facilitated the accurate estimation of survival times for cancer patients. 

Furthermore, Wang et al. \cite{wang2023penalized} introduced a survival analysis model in their research, utilising the Cox model in conjunction with XGBoost and an Elastic-net penalty to forecast survival. The authors assert that their approach is more flexible when dealing with survival data. Nevertheless, it is important to note that the models were trained using simulated data and then evaluated using four real datasets: the Veteran Administration lung cancer study, Stanford heart transplant data, Mayo Clinic primary biliary cirrhosis data, and MIMI-III from Beth Israel Deaconess Medical Centre.  }

Survival analysis has become increasingly popular across industries, including cybersecurity. For instance, to detect the insider threat actor \cite{alhajjar2021survival}, or to determine the level of exploitability risk from vulnerabilities \cite{roumani2021patching}. Survival analysis problems can be divided into three statistical methods \cite{wang2019machine}: non-parametric, semi-parametric, and parametric. Non-parametric methods are more efficient when no theoretical distribution is known and produce inaccurate results. On the other hand, the semi-parametric techniques do not require knowing the distribution of the survival times; however, the results may be more accurate as they are challenging to interpret. Lastly, the parametric methods provide accurate results and are easily interpreted when the distribution function is known. 

The Cox model is the most commonly used regression approach for survival analysis in semi-parametric methods. This model is built on the assumption of proportional hazards and utilises partial likelihood to estimate parameters. Despite not requiring the specification of the event time distribution, the attributes are assumed to influence the outcome exponentially. Parametric methods offer an alternative to Cox-based models (semi-parametric) and are also widely used to predict the time to an event of interest. 

Parametric survival models tend to produce estimates consistent with the theoretical survival distribution. The most commonly used distributions in parametric survival analysis are normal, exponential, Weibull, logistic, log-logistic, and log-normal. If all the survival times of the instances in the dataset are known and follow these distributions, the model is called a linear survival regression. Moreover, if the logarithm of the survival times follows these distributions, then the Accelerated Failure Time (AFT) model can be used to analyse the problem. 

{Unlike the Cox model, the AFT model is built on the assumption of a covariate that can accelerate or decelerate the time of an event (e.g. death). This model hence utilises the risk of death \cite{cleophas2023accelerated}. The hazard of death (used in the Cox model) is the ratio obtained between deaths and non-deaths (deaths/non-deaths), while the risk of death is the ratio of deaths and the entire population when the study began (deaths/all-population).} Another difference between hazard-based and death-based ratios is the type of results. While the hazard ratio returns values between zero and infinity, the death ratio returns a value between zero and one (0-100\%). 

\subsection{Survival Analysis to Predict Time to Fix for IoT}\label{closest}

Farris K. et al. \cite{farris2017vulnerability} explore using the Cox proportional hazards regression model to predict survival rates of software vulnerabilities. Their research analyzes trends in vulnerability survival rates, considering factors such as severity level, mission-critical status, and operating system type. A key contribution of the study is investigating the hazard rate associated with vulnerabilities’ time-to-remediation through survival analysis probability estimates and the Cox model. The dataset used in the study was derived from a Nessus server scan of 2,000 machines at a security operations centre over 12 months. Vulnerability remediation was defined as the event marking the “death” of a vulnerability, which occurs when a fix is applied. However, the dataset raises specific concerns. It was generated by a vulnerability management server, which could itself contain vulnerabilities, potentially influencing the results. Additionally, the study considers the date the server detects a vulnerability as the date of its discovery. Since the server only detects vulnerabilities when scanning a device, the data might not accurately represent the discovery timeline, introducing potential biases.

MITRE assigns vulnerabilities upon request, and the date of their release does not necessarily correspond to the exact year they were discovered. To address this, vulnerabilities are supplemented with related information that can help estimate their discovery date. Additionally, incorporating vulnerability severity into the analysis did not contribute meaningfully to the predictions. The study's calculation of vulnerability “death” raises concerns, as it is based on the time a vulnerability is detected as active on a host rather than when a fix is released. Moreover, the authors employed the Cox proportional hazards regression model to estimate the likelihood of a vulnerability being fixed within a given time frame. However, this approach primarily focuses on predicting the probability of remediation after a set period without accounting for vulnerabilities that were never fixed. In survival analysis, data corresponding to unresolved cases or extending beyond the observation period are typically classified as censored. The study did not consider this critical attribute, potentially affecting the robustness of the analysis.

Using survival analysis and the Cox regression model, Yaman R. \cite{roumani2020examining} explored the influence of previously unexplored factors on the patch release time of zero-day vulnerabilities. The dataset for the study was compiled from the National Vulnerabilities Database (NVD) and the Zero-day Initiative (ZDI). The primary event of interest was the feature “Patch-Release-Time,” with the Cox regression model used to calculate hazard ratios, indicating the impact of specific factors on the time to patch. The ZDI specifies a 120-day limit for vendors to address reported zero-day vulnerabilities. The study used this threshold to classify patches as timely; patches exceeding this limit were categorised as late and labelled as censored in the analysis. In this context, a zero-day vulnerability refers to a vulnerability identified and reported to the vendor, remaining undisclosed to the public until the vendor officially announces it. The study relies on ZDI as the sole data provider. As an organisation managing bounty programs for software vendors, ZDI maintains records of zero-day vulnerabilities and their eventual public disclosure. While this dataset is valuable, its scope is limited to vulnerabilities reported through ZDI’s programs.

The authors conducted a sub-analysis using the weakness type (CWE) to classify the vulnerabilities in this implementation. However, the authors looking to balance their dataset only used the six most prominent weaknesses out of 1,356 \footnote{https://cwe.mitre.org/data/definitions/1000.html}. Through this implementation, the authors tried to find the level of influence certain factors have over the time to release for zero-day software patches. However, since the data was provided by only one source, this became a problem that was left as a topic for future research. Lastly, the dataset was complemented with information from the NVD; however, the researchers found many inconsistencies and no other source available that could be used as a complementary data provider, forcing the study to use only the NVD and ZDI as data sources.

Othmane et al. \cite{ben2017time} conducted a study to identify the key factors influencing the time required to resolve security issues in the SAP software development process. The study utilised three regression models—linear regression, recursive partitioning, and neural network regression—to predict issue resolution times. Data for the analysis was derived from three datasets specific to SAP’s development processes. However, the reliance on a single data source and the equal treatment of issues and vulnerabilities represent key limitations. The study aimed to uncover the primary factors affecting the time to release fixes for security issues and bugs and predict the resolution times. Despite these efforts, the results showed no significant performance differences among the three regression models, and no dominant model emerged. The authors acknowledged this limitation, citing their inability to identify a suitable regression approach. Additionally, the study concluded that the “vulnerability type” feature—grouped into 511 categories—had no significant impact on issue resolution time. This conclusion, however, applies specifically to the SAP development platform. It is important to note that SAP’s development process involves two distinct phases: one within the development teams for reporting bugs and another where security analysts identify vulnerabilities in released products. This dual-phase approach, common among software corporations, complicates generalisations. The limited scope of the SAP-specific vulnerability classification process, which focuses on a single product, results in fewer recorded attacks, weaknesses, and vulnerability types. Consequently, efforts to generalise these findings will likely face challenges and produce suboptimal evaluation results. The authors acknowledged these constraints in the limitations and future work sections of their study.

We have summarised the comparison between our proposal and the main characteristics of the key related work to ours in terms of time to fix prediction using survival analysis in Table \ref{table:rw}.

\begin{table*}
\caption{Comparison of the features of our work with those in the related literature, and have created a breakdown to outline the differences.}
\label{table:rw}
\centering
\begin{tabular}
%Reference   |Prediction  |Approach used |DatasetSources |Generalisable?
{|p{2cm}   |p{2.7cm}    |p{2.7cm}      |p{2.3cm}       |p{2.5cm}|}
\hline
\textbf{Reference} &\textbf{Predicts} &\textbf{Approach used} &\textbf{Dataset Sources}&\textbf{Can it be generalisable}\\
\hline 
Farris K. et al. \cite{farris2017vulnerability} &Predict the lower survival rates for software vulnerabilities &Survival Analysis through Cox Proportional Hazard Regression. 
  &Nessus vulnerability analysis from 2000 computers &The authors consider it challenging to generalise this proposal.\\
\hline
Yaman R. \cite{roumani2020examining} &Predicts the hazard ratio for factors that affect patch time to release in Zero-day vulnerabilities. &Survival Analysis through Cox Proportional Hazard Regression.
  &NVD and ZDI &The authors concluded this proposal as not generalisable.\\
\hline
Othmane et al.\cite{ben2017time} &Predicts the time to fix for SAP bugs and vulnerabilities.            &Linear Regression      
 &SAP development. &The authors concluded this proposal as not generalisable.\\
\hline
Our solution &Predicts the time to fix IoT vulnerabilities.            &Survival Analysis through Accelerated Failure Time.      
 &MITRE, NVD, VulDB, and Twitter. &This approach is generalisable and transportable.\\
\hline
\end{tabular}
\end{table*}

\begin{table*}
\caption{Breakdown of the features provided by each source: MITRE (SRC-1), NIST (RC-2), VulDB (SRC-3), and Twitter (SRC-4).}
\label{table:table1_DS_Sources}
\centering
\begin{tabular}
{|p{1cm}       |p{2.5cm}     |p{1.8cm}      |p{1.3cm}    |p{5.5cm}|}
\hline
\textbf{Source}& \textbf{Feature Name}  &\textbf{Data Type} &\textbf{Unique Values} &\textbf{Details}\\
\hline 
SRC-1 &CVEID & Continuous            & 1,022 (21-02/2023)     &Vulnerability identifier.\\
\hline
SRC-2 &CVSS & Categorical            & 100           &A real number range 0 to 10.\\
\hline
SRC-2 &Weakness ID (CWEID) & Categorical & 926     &Weakness identifier\\
\hline
SRC-3 &Title & Categorical            & 885     &Vulnerability Title\\
\hline
SRC-3 &Summary & Categorical          & 271     &Vulnerability Summary\\
\hline
SRC-3 &Affected & Categorical         & 871     &Affected products description\\
\hline
SRC-3 &Vulnerability & Categorical    & 73     &Vulnerability details identifier.\\
\hline
SRC-3 &Impact & Categorical           & 19     &Attack impact\\
\hline
SRC-3 &Exploit & Categorical          & 91     &Exploit description\\
\hline
SRC-3 &Countermeasure & Categorical   & 194     &Suggested countermeasure description\\
\hline
SRC-3 &Sources & Categorical          & 735     &Data sources description\\
\hline
SRC-3 &Vendor & Categorical           & 77     &Vendor name of affected product\\
\hline
SRC-3 &Product Name & Categorical     & 333     &Name of affected product\\
\hline
SRC-3 &Component & Categorical        & 360     &Name of affected component\\
\hline
SRC-3 &Version & Categorical          & 392     &Affected product version\\
\hline
SRC-3 &Risk & Categorical             & 3     &Vulnerability Risk (Low, Medium, High)\\
\hline
SRC-3 &Class & Categorical            & 72     &Vulnerability Class Name\\
\hline
SRC-3 &Attck & Categorical            & 21     &Attck Identifier from MITRE Att\&ck\\
\hline
SRC-3 &CVE Assigned &Date             & 235  &Date of the CVE assignment\\
\hline
SRC-3 &VulDB Base Score &Categorical  & 100     &CVSSv3 Base Score calculation by VulDB\\
\hline
SRC-3&Vulnerability Found&Date        & 144     &Timestamp for reported vulnerability found\\
\hline
SRC-3 &Advisory Disclosed&Date        & 326     &Timestamp for reported vulnerability Disclosed by Advisory\\
\hline
SRC-3 &CVE Reserved & Date            & 235     &Timestamp CVE Reserved\\
\hline
SRC-3 &NVD Disclosed & Date           & 215     &Timestamp of Vulnerability Disclosure by NVD\\
\hline
SRC-3&VulDB Entry Created&Date        & 171     &Timestamp of Vulnerability Entry Creation by VulDB\\
\hline
SRC-3&VulDB Entry Updated &Date       & 1022     &Timestamp of Vulnerability Entry Last Updated by VulDB\\
\hline
\end{tabular}
\end{table*}

\begin{table*}
\caption{Breakdown of the features provided by each source: MITRE (SRC-1), NIST (RC-2), VulDB (SRC-3), and Twitter (SRC-4)...(Part two)}
\label{table:table2_DS_Sources}
\centering
\begin{tabular}
{|p{1cm}       |p{2.5cm}     |p{1.8cm}      |p{1.3cm}    |p{5.5cm}|}
\hline
\textbf{Source}& \textbf{Feature Name}  &\textbf{Data Type} &\textbf{Unique Values} &\textbf{Details}\\
\hline
SRC-3&VulDB Entry Updated &Date       & 1022     &Timestamp of Vulnerability Entry Last Updated by VulDB\\
\hline
SRC-3&Advisory Confirmation&Date      & 3     &Advisory Confirmation Type (Confirmed, Not Defined, Uncorroborated \\
\hline
SRC-3 &Exploit Publicity & Categorical& 3     &Publicity of the exploit (Public, Private, NA)\\
\hline
SRC-3&Exploit Availability&Categorical& 3     &Availability of Exploit (1=Yes, 0=No, NA)\\
\hline
SRC-3 &Price 0day & Categorical       & 4     &Known or Estimated 0-day Price of the Exploit\\
\hline
SRC-3 &Price Today & Categorical      & 4     &Known or Estimated Price of the Exploit as of Today (Daily Updated)\\
\hline
SRC-3 &EPSS Score & Categorical       & 536     &Current Value of the Exploit Prediction Scoring System\\
\hline
SRC-3 &EPSS Percentile & Categorical  & 653     &Percentile of CVE within Current EPSS\\
\hline
SRC-3&Countermeasure&Categorical      & 4     &Generic Remediation Level Description\\
\hline
SRC-3&Countermeasure Name&Categorical & 8     &Name of the Suggested Countermeasure\\
\hline
SRC-3 &Upgrade Version & Categorical  & 132     &First Known Unaffected Version(s)\\
\hline
SRC-4 &Tweets & Categorical           & 54     &Tweet Count of a CVE\\
\hline
SRC-4 &First Tweet Date & Date        & 165     &Timestamp of the First Tweet\\
\hline
SRC-4 &Retweet Count & Categorical    & 44     &Retweets Average\\
\hline
SRC-4 &Reply Count & Categorical      & 21     &Reply-Count Average\\
\hline
SRC-4 &Like Count & Categorical       & 47     &Like-Count Average\\
\hline
SRC-4 &Quote Count & Categorical      & 19     &Quote-Count Average\\
\hline
SRC-4 &Impression Count & Categorical & 18     &Impression-Count Average\\
\hline
\end{tabular}
\end{table*}

\section{Time to fix prediction for IoT devices through Survival Analysis and Regression Models}\label{mainsection}

In this section, we provide details on each element that encompasses our time to fix prediction framework. We start with the conceptual analysis of the datasets in Section \ref{datasets}. Then, Section \ref{framework} expands on the proposed framework, while  Section \ref{em4aft} discusses the evaluation metrics we selected for our model. 

\subsection{Dataset}\label{datasets}
%{A research study conducted by John Smith \cite{smith2021addressing} highlighted the importance of maintaining data quality in a dataset. The study emphasised the significance of addressing challenges related to accuracy and completeness throughout data collection, processing, cleaning, and managing missing data. In the dataset constructed within this paper, meticulous attention was given to each stage to mitigate potential issues and uphold data integrity. During data collection, rigorous efforts were made to verify that the selected sources could furnish 100\% of every feature. Subsequently, a single individual meticulously
%processed and cleaned data without external intervention. Notably, an outstanding characteristic of our dataset is its absence of missing data.}
    
Our dataset was created by collecting vulnerability information on IoT devices from four sources: MITRE\footnote{https://cve.mitre.org/}, NIST NVD\footnote{https://nvd.nist.gov/}, VulDB\footnote{https://vuldb.com/}, and Twitter\footnote{https://twitter.com}. MITRE provided information about common vulnerabilities and exposures (CVE) and their corresponding release dates. NIST delivers additional data to each CVE, such as the common vulnerability risk score (CVSS) and the common weaknesses enumeration (CWE). VulDB was used to obtain extensive information on the vulnerability (CVE), including risk scores (CVSS), exploits, attacks, weaknesses (CWEs), remediation actions, and dates corresponding to each element. Lastly, Twitter was used to identify social network trends related to each published vulnerability identifier (CVEID) and extracted only the trend counts provided by the platform. Table \ref{table:table1_DS_Sources} provides a detailed breakdown of the features we extracted from each source.
\newline

We used the MITRE and NIST vulnerabilities databases (NIST NVD) to gather vulnerabilities. We filtered out records not related to hardware by identifying the type of affected product using the product code in the NVD (\textit{o} for operating system, \textit{s} for software, \textit{a} for application, and \textit{h} for hardware). We utilised the publication date of CVE as a reference to determine the time to fix for each vulnerability. However, we came across numerous records that resulted in a zero time to fix. This indicates that the date of the fix being published is the same as that of CVE. There could be various reasons behind this, and one of the most common reasons we found is the manufacturer's strategy, which involves publishing CVEs only after developing the fix. Due to this biased process, we conducted an in-depth analysis of the CVE data. We discovered that NVD labels the references attached to each CVE, and we found it convenient that one of the labels refers to PATCH (Vendor Advisory, Patch, Third Party Advisory). After developing Java code scripts, the dataset underwent filtration procedures intending to retain only those records that were accurately labelled as PATCH. 

To accurately gauge the length of time to fix, we thoroughly investigated the reference links associated with each CVE. We meticulously searched for relevant information regarding the date the fix was made available. Once we obtained this information, we compared it to the date the vulnerability had been initially detected. By subtracting the latter from the former, we could precisely calculate the time (time to fix) it had taken for the device's manufacturer to address the reported vulnerability (CVE). The calculated time to fix is what we will use as the predicted feature. 

Upon compiling all pertinent information, our final dataset consisted of 1,027 records. These records encompass many aspects, including vulnerabilities, exploits, remediation actions, risk levels, timelines, and Twitter trends regarding IoT devices. We thoroughly analysed this dataset to identify the optimal model for implementation. The dataset created for this project is made publicly available: \url{https://github.com/criveraalvarez/Time2FixPredictionDS/}.

\subsection{Framework Conception}\label{framework}
Based on our comprehensive dataset analysis and the time-based prediction feature, we have decided to refrain from employing deep learning models. Our reasoning stems from a recent research by Shwartz-Ziv and Armon \cite{Shwartz-ZivA22}, that comments that when considering using tabular datasets (similar to our own), the XGBoost model outperformed deep learning models across eleven different datasets. Despite the general dataset size being over seven thousand, the study found that deep learning models only performed well on the specific dataset they were designed for. XGBoost consistently performed well across all datasets.

We perform a regression task to accurately predict the time (in days) for a vulnerability to be fixed, which is essential for our time to fix feature. The prediction of the time to fix we seek can be achieved using Survival Analysis as demonstrated in \cite{faruk2018comparison}, which utilised the Accelerated Failure Time (AFT) model. To deploy the AFT regression approach, we confidently employ the extreme gradient boosting ensemble (XGBoost) model \cite{wang2023penalized, BarnwalCH22}. In addition to prior work, Barnwal et al. \cite{barnwal2022survival} conducted experiments involving integrating the Accelerated Failure Time (AFT) model with the XGBoost library. Their findings revealed that using the second-order partial derivative of the loss function in XGBoost accelerates convergence. Consequently, integrating the AFT model with XGBoost simplifies survival analysis, particularly in the context of handling extensive datasets.  

The Accelerated Failure Time (AFT) model is represented as:
\begin{equation} \label{AFT}
\begin{split}
\ln Y= \bigl \langle w,x \bigr \rangle + \sigma Z
\end{split}
\end{equation}
\newcommand{\R}{\mathbb{R}}

\noindent here, $\ln x$ is the natural logarithm of $Y$; $x$ is a vector in $\R^d$ that represents the features in our dataset; $w$ is a vector of $d$ coefficients (each corresponding to a feature); the dot product between w and x is represented by $\bigl \langle w,x \bigr \rangle$ in $\R^d$; $Y$ is the output feature (time to fix), and $Z$ is a random variable of a known probability distribution, such as Normal, logistic, or extreme distribution. And, $\sigma$ is a parameter that scales $Z$ size.

To make AFT work with gradient boosting models, we revise the Equation \ref{AFT} into the following:
\begin{equation} \label{AFT-GB}
\begin{split}
\ln Y=  \tau(x) + \sigma Z
\end{split}
\end{equation}

\noindent where, $\tau(x)$ represents a decision tree ensemble output, $x$ is the input, and $Z$ is a random variable. The XGBoost ensemble-tree model\footnote{https://xgboost.readthedocs.io} maximises the $log-likelihood$ by fitting the ensemble $\tau(x)$.

Following Barnwal et al.\cite{BarnwalCH22}, we utilise XGBoost, which defines the probability density function for $\mathcal{D}$ (the training data) as the product of probability densities $f_Y$ for each data point in its maximised form:
\begin{equation}\label{Likelihood}
\begin{split}
\ln L(\mathcal{D})=\sum^n_{i=1} \ln\mathbb{P}[Y_i=y_i]=\sum^n_{i=1}\ln f_Y(y_i) 
\end{split}
\end{equation}

In survival analysis, the feature $y_i$ may be unknown for certain data points; hence, it is considered censored. The revised version of the maximised likelihood function considering censored data points would be as follows:
\begin{equation}\label{Likelihoood_censor}
\begin{split}
\ln L(\mathcal{D})=\sum^n_{i=1}\ln f_Y(y_i)+\sum^n_{i=1} \ln(f_Y(\overline{y_i})-f_Y(\underline{y_i})) 
\end{split}
\end{equation}

\noindent where, $f_Y(\overline{y_i})$ and $f_Y(\underline{y_i})$ represent the upper and lower bounds for the feature $y_i$ respectively. 

In AFT, the loss function is based on a known Probability Density Function (PDF) and a Cumulative Distribution Function (CDF). 
\begin{equation}
\begin{split}
l_{AFT}(y,\tau(x)) = \Bigg\{ -\ln[f_Z(s(y)) \cdot \frac{1}{\sigma y}] \\
                            \text{if $y$ is not censored} \\
\end{split}
\end{equation}

\begin{equation}
\begin{split}
l_{AFT}(y,\tau(x)) = \Bigg\{ 
        -\ln[f_Z(s(\overline{y}))-f_Z(s(\underline{y}))] \\
        \text{if $y$ is censored with $y \epsilon [\underline{y},\overline{y}]$}
\end{split}
\end{equation}

\begin{table*}[h!]
\caption{The features are divided into six main groups relevant to the research question (Section \ref{litrev}).}
\label{table:groups}
\centering
\begin{tabular}
{|p{3cm} |p{10.2cm}|}
\hline
\textbf{Group}& \textbf{Features}  \\
\hline 
Basic Data & 'Vulnerability', 'Vendor', 'Device Name', 'Affected Products', 'Version' \\
\hline
VulDB Data & 'Affected Models', 'Impact', 'Exploit Existence', 'Countermeasures', 'Sources', 'Risk', 'Class', 'Attack', 'Vulnerability Found', 'Advisory Disclosed', 'CVE Reserved', 'NVD Disclosed', 'Vuldb Entry Created', 'Vuldb Entry Updated', 'Advisory Date', 'Advisory Confirmation', 'Exploit Availability', 'Exploit Publicity', 'Exploit Exploitability', 'Price0day', 'PriceToday', 'Epss Score', 'Epss Percentile', 'Remediation', 'Countermeasures Name', 'Upgrade Versions', 'Vuldb CVSS BS'\\
\hline 
NIST Data &'CVE-ID', 'CVE Assigned', 'CVE Published', 'NVD CVSS BS', 'CWE-ID'\\
\hline
S2V Title Data &'S2V-Title Sum'\\
\hline
S2V Summary Data &'S2V-Summary Sum'\\
\hline
Twitter Trend Data &'Tweets', 'Date First Tweet', 'Retweet Average', 'Reply Average', 'Like Average', 'Quote Average', 'Impression Average'\\
\hline
\end{tabular}
\end{table*}

XGBoost offers three function options: Normal, Logistic, and Extreme. Our testing showed that the Normal function produced better results.

Survival analysis utilises a tuple as the predicted feature $Y$. This tuple comprises the $time$ in days where the fix is observed to be delivered for any IoT vulnerability. The $event$ specifies whether the fix is provided or not. Survival analysis considers it censored when an event is not observed. In the context of our dataset, it refers to an unknown provision of a fix (NA time to fix). 

It is noteworthy to say that our dataset does not include the features "event" and "time2fix." To enable the XGBoost model to operate effectively, we incorporated these features. The "event" attribute is derived from a function that evaluates the "Remediation" feature of the dataset. In one instance, the outcome indicates the availability of a fix, while in all other scenarios, it pertains to the absence of a remedy. The attribute "time2fix" is determined through a comprehensive analysis of the references associated with each CVE. To accomplish this, we employed a script that automatically combed through the references (URLs) to locate the date the CVE was discovered and when a fix/no-fix was issued.

In survival analysis, there are three types of censorship: right, left, and interval. If an IoT device's vulnerability receives a fix after several months or even years, it is known as right-censored. Conversely, if a vulnerability gets a fix before the average time, it is called left-censored. In cases where a vulnerability is fixed within an unspecified timeframe but falls between the average times, it is referred to as interval-censored. 
{Our dataset analysis reveals that individual records feature unique time to fix values, which can fall below, above, or match the average. Hence, the most optimal strategy would involve the utilisation of interval censoring. }

\begin{table}[h!]
\caption{List of XGBoost parameters and their best-performing values.}
\label{table:parameters}
\centering
\begin{tabular}
{|p{4cm} |p{2cm}|}
\hline
\textbf{Parameter}& \textbf{Value}  \\
\hline 
Learning Rate & 0.0002\\
\hline
Max depth & 8\\
\hline 
Sub-sample &0.5\\
\hline
Min child weight &50\\
\hline
Col sample by node &0.5\\
\hline
Lambda &0.01\\
\hline
Alpha &0.02\\
\hline
\end{tabular}
\end{table}

\subsection{Evaluation Metrics for an AFT model}\label{em4aft}
{The presence of censoring data requires specific approaches to evaluate the performance of the AFT model results. In the publication by Wang et al. \cite{wang2019machine}, the authors suggest the utilisation of the Concordance Index (C-Index), Brier Score, and Mean Absolute Error as recommended metrics. Among these metrics, the C-index stands out as it calculates the probability of concordance or the relative risk of an event occurring for various instances. This recommendation holds particular relevance in our case, whereby using these metrics would be pivotal in assessing our model’s performance.} Thence, the C-index provides the probability for time to fix observation and prediction results from the two instances, $(y_1,\hat{y}_1)$ and $(y_2,\hat{y}_2)$, The C-index will hence use the following equation. 

\begin{equation}\label{c-index}
\begin{split}
c=Pr(\hat{y}_1 > \hat{y}_2 | y_1 \geq y_2)
\end{split}
\end{equation}

Here, $y_1$/$y_2$ and $\hat{y}_1$/$\hat{y}_2$ represent the actual time to fix and the predicted value, respectively. 
Next, we will explore how to implement the survival analysis AFT with the XGBoost model and examine the resulting evaluation metrics.

\section{Experiments and Results}\label{ExperimentsandEvaluationResults}
We utilised the Colab platform, which runs on Python, to carry out the Survival Analysis AFT through the XGBoost \cite{chen2016xgboost, BarnwalCH22} model. Before feeding the XGBoost model with the dataset, we meticulously analysed it through various methods we explain next. 
\newline

In the data examination process (Tables \ref{table:table1_DS_Sources} and \ref{table:table2_DS_Sources}), we discovered that most feature values can be transformed into numerical values using the Frequency Label-Encoder function from Sklearn\footnote{https://scikit-learn.org}. We found two features of string-based data, namely vulnerability title and summary. It is not feasible to convert these to numeric values as the strings are unpredictable and vary in length. We could convert sentences into numeric vectors using a modified version of the word2vec method called sentence2vec\footnote{https://pypi.org/project/sent2vec/}. By summing up the vectors' values, we obtained two new features: S2V-Title and S2V-Summary. These features were then grouped into six categories related to our research question (Section \ref{litrev}) of predicting the time to fix for an IoT vulnerability. The distribution of features for each group can be found in Table \ref{table:groups}.

\begin{figure*}[t]
    \includegraphics[width=1\textwidth]{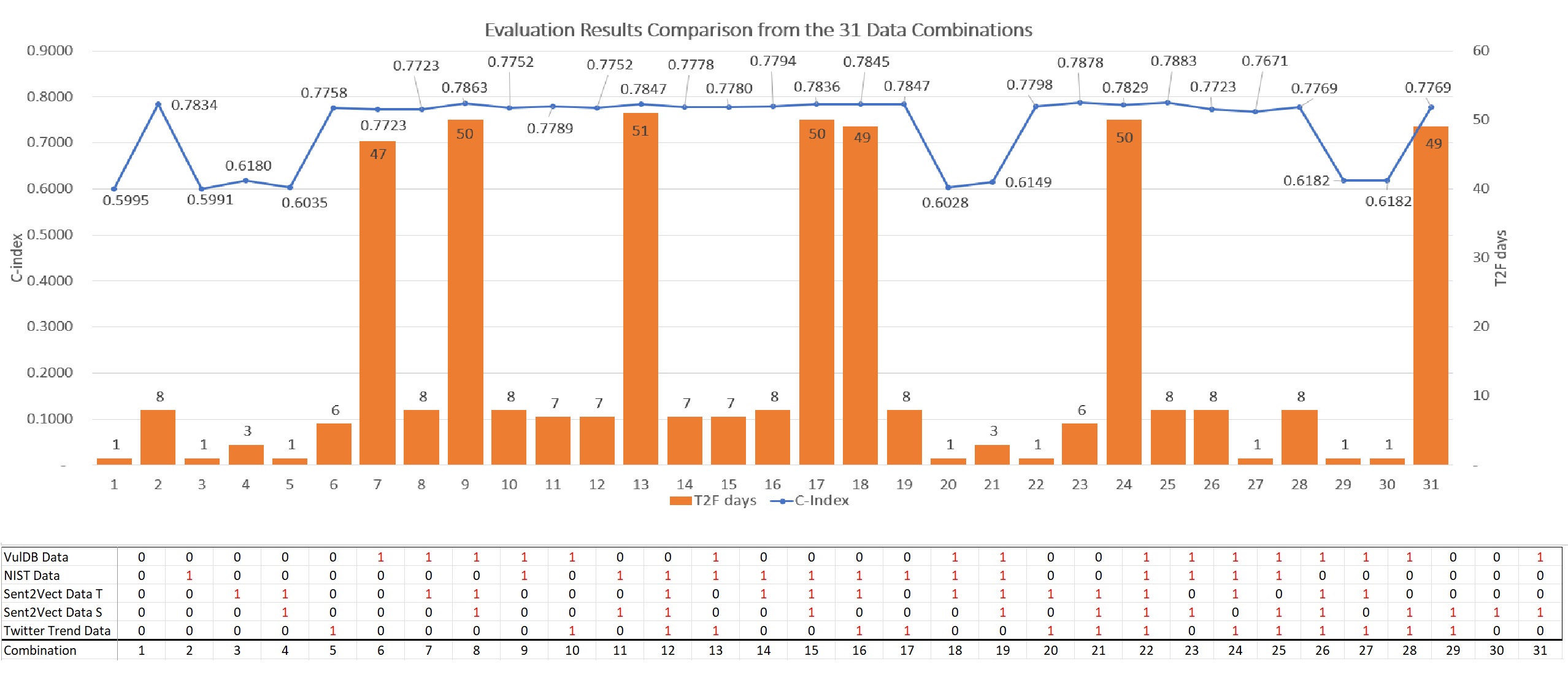}
    \centering
    \caption{C-Index and time to fix in days for all the 31 group combinations.}
    \label{fig:results31}
\end{figure*}

The XGBoost algorithm utilises a tuple to label data. This tuple includes the time $T$ and the event $E$, indicating whether a fix was implemented. To create it, we combine the time to fix as the time element and the binary values from the feature remediation, which describe the nature of the event.

\begin{figure}[t]
    \includegraphics[width=1\textwidth]{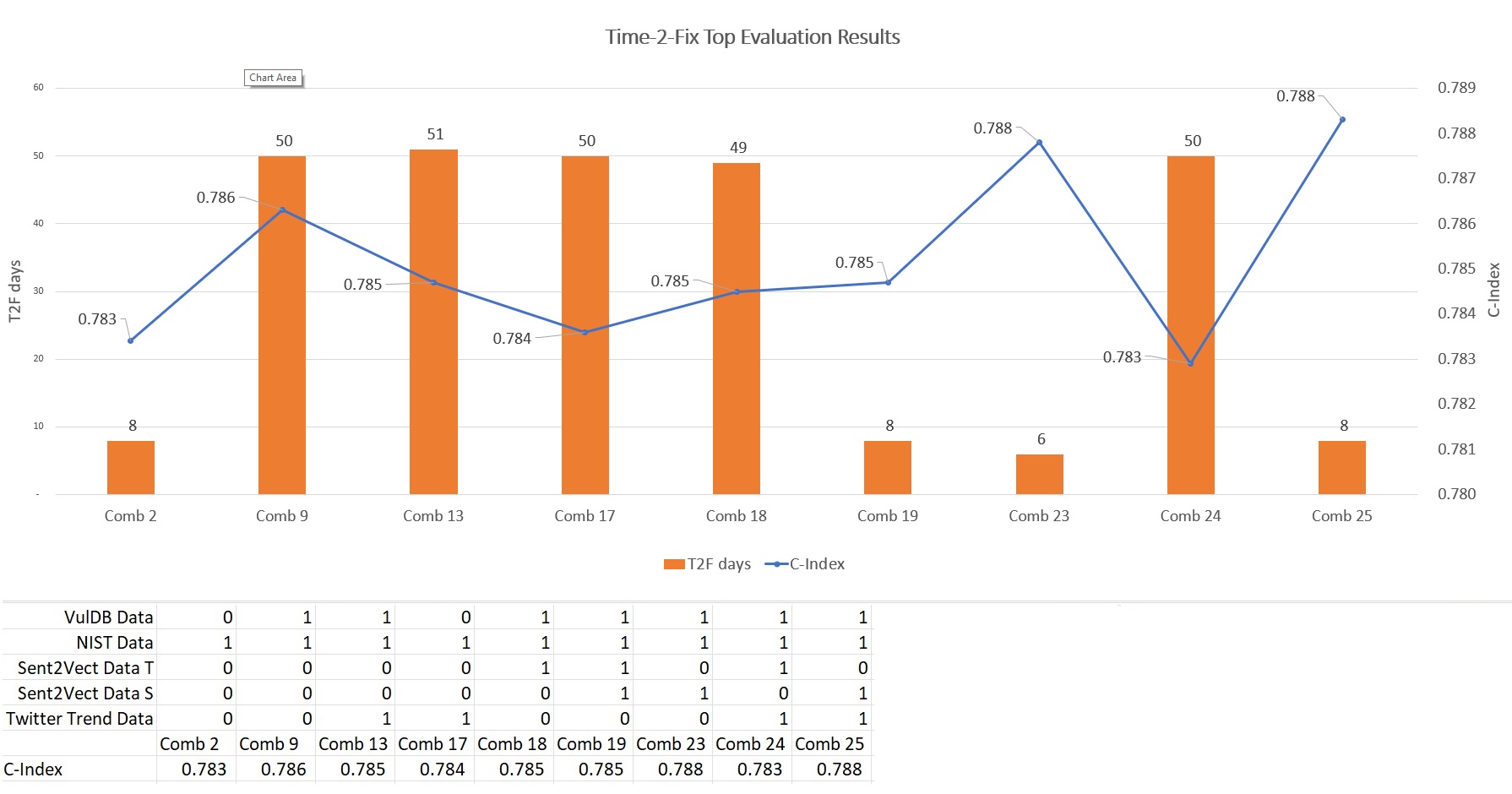}
    \centering
    \caption{C-Index and time to fix in days for the top 9 C-index group combinations.}
    \label{fig:topresult}
\end{figure}

{To identify the most optimal parameter values, we followed a similar methodology to that of Barnwal et al. \cite{BarnwalCH22}, where they compared default values, grid search, and random search. Initially, we utilized the default parameter values and refined them using the fine-tuning method (GridSearch) to identify the best-performing values. The selected parameters are presented in Table 1 for reference.} Additionally, we utilised the early stopping technique to prevent overfitting while training the model. As shown in Figure \ref{fig:results31}, we disclose the C-Index for each data combination and its corresponding predicted time to fix. It is important to note that we confidently default to the Basic Data group as the most straightforward combination. Therefore, we incorporate one distinct mix from the other groups, resulting in 31 available combinations. This includes the option that solely utilises the primary data.

In Figure \ref{fig:topresult}, we present a table with the combinations that yielded the highest C-Index values. The figure demonstrates that data from VulDB and NIST (CVE, CWE, and CVSS) and the groups using summed vectors from the title and summary can predict the time to fix for an IoT vulnerability. Furthermore, our findings indicate that incorporating data from Twitter trends offers a minimal advantage to the model. We plot (Figure \ref{fig:error}) the error function (negative logarithm likelihood) from training and validation.

When evaluating survival analysis models, utilising the C-Index is the recommended option; however, it is essential to understand that it produces results on a scale from 0 to 1, but this does not necessarily equate to a traditional probability outcome. The C-Index values will differ depending on the dataset and model used \cite{LiuFYDZZ21,brentnall2018use,heller2016estimating}. Lastly, to ensure the implementation generates dependable results, it's crucial to observe the behaviour of the error function \cite{LiuFYDZZ21}. 

\begin{figure}[t]
    \includegraphics[width=1\textwidth]{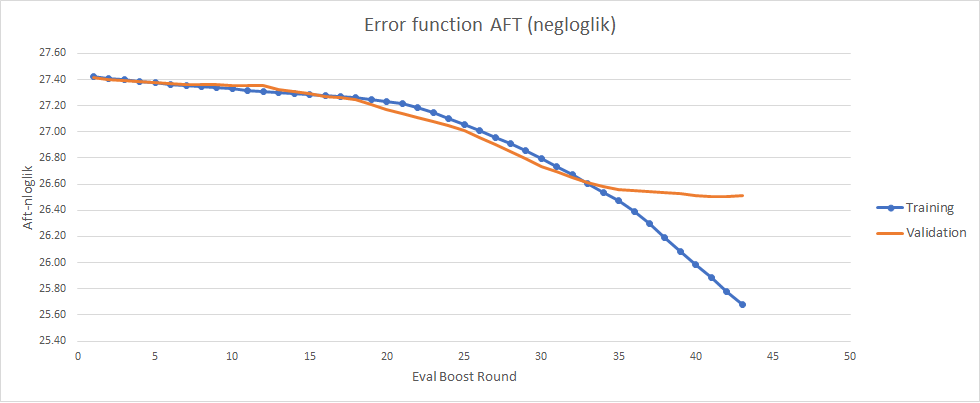}
    \centering
    \caption{Train and Validation Error Function (negloglik) Results from combination No. 25 that provided the highest C-Index.}
    \label{fig:error}
\end{figure}

We compared evaluation results using the C-Index and error function (Figure \ref{fig:error}). Our analysis of the error function at iteration 35 for the model with the highest C-Index values indicated that the model is stable during validation, demonstrating good generalisation. This suggests that the model can effectively fit the training dataset. However, we also observed an accelerated decrease in the loss function of the training dataset after this iteration, indicating overfitting. Our findings determined that the XGBoost model (Survival Analysis AFT) can predict time to fix for an IoT vulnerability using data from VulDB and NVD, along with vectorised summation from Vulnerability Title and Summary. Ten feature variations (shown in Figure \ref{fig:topresult}) can be used. Incorporating Twitter CVE trends into the model provides minimal additional information and may be optional for future datasets.  

\section{Conclusion and Future Work}\label{Conclusion}

We proposed a survival analysis framework to predict the time required to fix vulnerabilities (time to fix) in IoT devices, enabling system administrators to manage patching processes more effectively. We employed the accelerated failure time (AFT) model instead of the commonly used Cox proportional hazard model to predict time to fix for IoT vulnerabilities. This approach was implemented using the XGBoost machine learning model and based on a new dataset we created, which includes information on IoT devices, detected vulnerabilities, and social network trends from Twitter.

We trained the XGBoost model with different combinations of features to test our hypotheses regarding the predictive power of various data sources. The evaluation results demonstrated that using public IoT device vulnerability and vulnerability-specific information effectively predicts time to fix, achieving satisfactory C-index levels. However, incorporating Twitter social trends did not improve the model performance in our study.

Our study introduces a novel methodology for predicting time to fix in IoT systems and opens several avenues for future research:

\begin{itemize}
\item \textbf{Dataset Enhancement}: Expanding the dataset with additional features and more records could enhance the model’s predictive capabilities, as our analysis highlights the significant impact of individual features on the results.
\item \textbf{Automated Data Gathering}: Automating the content validation process for new features could improve data collection and analysis efficiency and scalability.
\item \textbf{Alternative AI Models}: Exploring other AI models, such as deep learning architectures like long short-term memory (LSTM) networks, may improve predictive performance, as suggested by prior studies.
\item \textbf{Extending the Methodology}: Future research could apply our time to fix methodology to predict not only fix times but also the emergence of vulnerabilities in IoT systems and related domains.
\end{itemize}

Addressing these areas can enhance the predictive accuracy of time to fix models and improve vulnerability management in IoT environments.

\bibliographystyle{ACM-Reference-Format}
\bibliography{references}

\end{document}